\documentclass[9pt]{pnas-new}
\usepackage[font=small,labelfont=bf,justification=justified,calcwidth =.99\linewidth]{caption}
\usepackage{xcolor}
\usepackage{amsmath,amssymb}
\usepackage[separate-uncertainty,detect-family=true,mode=math]{siunitx}
\usepackage{bm}

\renewcommand*{\vec}[1]{\bm{#1}}

\newcommand*{\norm}[1]{\lVert #1 \rVert}

\usepackage{natbib}

\makeatletter
\renewcommand{\maketitle}{\noindent\textbf{\Huge\@title}\bigskip\par{\@author}\bigskip\par}
\makeatother

\title{Swimmers at Interfaces Enhance Interfacial  Transport}
\author{Jiayi Deng, Mehdi Molaei, Nicholas G. Chisholm and Kathleen J. Stebe}
\correspondingauthor{Kathleen J. Stebe\\E-mail: kstebe@seas.upenn.edu}

\begin{document}

\maketitle
The behavior of fluid interfaces far from equilibrium plays central roles in nature and in industry. Active swimmers trapped at interfaces can alter transport at fluid boundaries with far reaching implications. Swimmers can become trapped at interfaces in diverse configurations and swim persistently in these surface adhered states. The self-propelled motion of bacteria makes them ideal model swimmers to understand such effects. We have recently characterized the swimming of interfacially-trapped \textit{Pseudomonas aeruginosa} PA01 moving in pusher mode. The swimmers adsorb at the interface with pinned contact lines, which fix the angle of the cell body at the interface and constrain their motion. Thus, most interfacially-trapped bacteria swim along circular paths. Fluid interfaces form incompressible two-dimensional layers, altering leading order interfacial flows generated by the swimmers from those in bulk. In our previous work, we have visualized the interfacial flow around a pusher bacterium and described the flow field using two dipolar hydrodynamic modes; one stresslet mode whose symmetries differ from those in bulk, and another bulk mode unique to incompressible fluid interfaces. Based on this understanding, swimmers-induced tracer displacements and swimmer-swimmer pair interactions are
explored using analysis and experiment. The settings in which multiple interfacial swimmers with circular motion can significantly enhance interfacial transport of tracers or promote mixing of other swimmers on the interface are identified through simulations and compared to experiment. This study identifies important factors of general interest regarding swimmers on or near fluid boundaries,
and in the design of biomimetic swimmers to enhance transport at interfaces.

%

\section{Introduction}
Bacteria have long been studied as prototypical active colloids whose self-propulsion and interaction generate exciting collective behaviors. Swimming bacteria share a common and simple machinery, with rotating motors in their cell envelope that are coupled to the flagella, enabling bacterial self-propulsion \cite{terashima_chapter_2008}. Hydrodynamic interactions and the chaotic nature of bacterial swimming give rise to intriguing non-equilibrium phenomena in active suspensions, including enhanced diffusion \cite{wu_particle_2000,kanazawa_loopy_2020,mino_enhanced_2011, underhill_diffusion_2008,kasyap_hydrodynamic_2014, pushkin_fluid_2013,mino_induced_2013}, long-range correlations in velocity and orientation fields \cite{kurtuldu_enhancement_2011,nambiar_enhanced_2021,miles_active_2019, mathijssen_nutrient_2018, belan_pair_2019, skultety_swimming_2020, krishnamurthy_collective_2015} and active phase separation \cite{thutupalli_flow-induced_2018,cates_motility-induced_2015,zhang_active_2021,alert_active_2022}.
These findings have inspired biomimetic active colloid systems designed to recapitulate bacteria’s swimming and collective behavior \cite{kokot_active_2017,spellings_shape_2015,sokolov_swimming_2010,shklarsh_smart_2011,di_leonardo_bacterial_2010}.

Interactions between swimmers and passive tracers in liquid suspension and near solid surfaces reveal enhanced diffusion and non-Gaussian statistics that differ from thermally driven Brownian motion \cite{wu_particle_2000}.
Tracers advected by swimmers moving along straight paths in bulk and near solid boundaries have been studied extensively to understand the enhanced transport by active colloids \cite{mathijssen_tracer_2015,mino_induced_2013,mino_enhanced_2011,morozov_enhanced_2014,pushkin_fluid_2013, kanazawa_loopy_2020, kasyap_hydrodynamic_2014,pushkin_fluid_2013-1, deng_interfacial_2023}. 
While the role of solid boundaries in altering swimmer-colloid and swimmer-swimmer interactions is well appreciated, the role of fluid interfaces in altering such interactions has not been addressed. 

\begin{figure}[t]
\centering
\includegraphics[width=.76\linewidth]{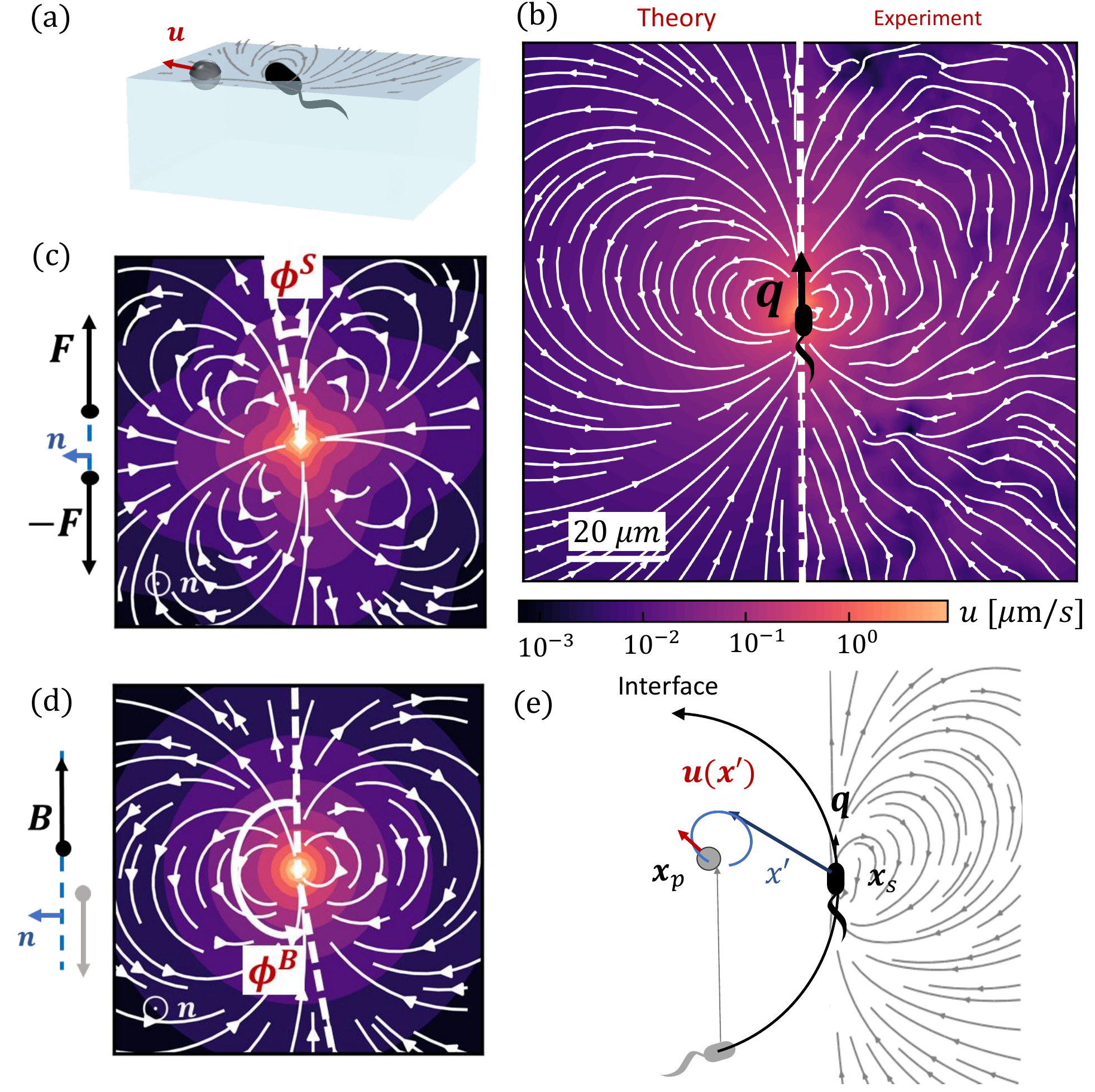}
\caption{\textbf{Dipolar flow generated by a pusher bacterium on a fluid interface} (a) Schematic of an interfacially trapped bacterium interacting with a tracer particle. (b) Measured flow field generated by an ensemble of pusher bacteria in origin moving in the \(y\) direction (left panel). 
Theoretical fit of the flow field generated by a pusher on the interface (right panel). Streamlines indicate the local direction of the flow, and the color scheme indicates its magnitude. The flow field is reprinted from \cite{deng_interfacial_2023}.
The dipolar flow consists of (c) \textbf{S} and (d) \textbf{B} modes. The strength, positions, and orientations for both modes are fitted simultaneously to the observed field. Specifically, this entails fitting the strength $S$, and orientation $\phi^S$ of the \textbf{S} mode, the strength $B$ and orientation $\phi^B$ of the \textbf{B} mode and the common position $\delta y$ of both modes using the non-linear least squares fitting method in python (scipy.optimize.curvefit). 
(e) Schematic of a tracer particle entrained by a bacterium swimming along a circular path.}
\label{fig1}
\end{figure}

We have previously studied swimming behaviors of \textit{Pseudomonas aeruginosa} PA01 at aqueous-hexadecane interfaces. These monotrichous bacteria move in pusher or puller modes by reversing the sense of rotation of their single flagellum. The bacteria can swim adjacent to the interface, or they can adsorb with cell bodies spanning the interface and swim in an adhered state, a phenomenon unique to fluid interfaces.  
In this surface-adhered state, the bacteria cell bodies become trapped at fluid interfaces with pinned contact lines, constraining their swimming behavior (Fig. \ref{fig1}a).  Furthermore, surfactant adsorption to the fluid interfaces gives rise to complex surface stresses. For example, surfactants can generate  Marangoni stresses that render the interface incompressible and alter swimmer motion. 

Interface-associated bacteria have complex trajectories that differ from those in bulk or near solid surfaces \cite{lopez_dynamics_2014, lauga_swimming_2006}. Interfacially trapped motile bacteria swim preponderantly in curly or circular paths with curvatures ranging from $0.1-$\SI{1.0}{\micro m^{-1}}. This circular swimming generates a ‘self-caging’ plateau in their mean square displacements (MSD). 
Weak displacements of their centers of rotation owing to active diffusion processes decorrelate their positions at long lag times \cite{deng_motile_2020} allowing them to move diffusively at the interface. 
The bacteria's active diffusivities are attributed to diverse athermal stochastic or noisy processes including fluctuations in flagellar rotation, switching between pusher and puller modes and interactions with other swimmers \cite{deng_interfacial_2023, grosmann_diffusion_2016}. 
 
We have measured the ensemble-averaged interfacial flow field generated by a PA01 bacterium moving in a pusher mode using correlated displacement velocimetry \cite{molaei_interfacial_2021, deng_interfacial_2023} (Fig. \ref{fig1}a). The bacteria and passive tracer colloids were trapped at interfaces of hexadecane and aqueous suspensions of bacteria in TRIS-based motility medium buffer; details are given in  \cite{deng_interfacial_2023}.
This flow field features unexpected asymmetries that do not arise for their bulk-fluid counterparts (see Fig. \ref{fig1}b). Analysis reveals that the interfacial velocity field can be decomposed into two dipolar hydrodynamic modes associated with interface incompressibility, an interfacial stresslet (\textbf{S} mode) corresponding to a parallel force dipole on the interface (Fig. \ref{fig1}c), and a second mode (\textbf{B} mode) generated from off-interface forcing by the flagellum immersed in the bulk fluid beneath the interfacial plane that is balanced by Marangoni stresses (Fig. \ref{fig1}d).
The relative importance of these modes is determined by the cell bodies’ trapped configurations \cite{deng_interfacial_2023}. 

In this study, we exploit these findings to analyze the impact of interfacially-trapped bacteria as model swimmers on interfacial transport (see schematics in Fig. \ref{fig1}e). 
The measurement of the flow fields and their decomposition into leading order hydrodynamic modes allows analytical prediction of tracer advection via hydrodynamic interactions with swimmers, and of pair interactions between swimmers. 
 To appreciate the importance of the circular trajectories on bacteria-tracer interactions, we consider swimmers that follow either straight or circular trajectories and calculate the corresponding trajectories of tracer particles. We find that circular swimmers at interfaces generate tracer displacements in which tracers also move in closed loops unless the swimmers and tracer particles are in close proximity.  
 We also study the manner in which neighboring swimmers interact, and find circular swimmers in close proximity can reorient and generate net displacements of their neighbors.
Interactions between a tracer particle and multiple swimmers are then studied to understand the impact of hydrodynamic interactions on tracer diffusion processes. For scant swimmers and low active noise, tracers have self-caged displacements owing to their loopy displacement trajectories. We find that higher swimmer concentration and active noise allow swimmers to break the caging effect, and therefore to further enhance interfacial mixing.
To further understand the role of swimmers in generating active noise in the interface, we study hydrodynamic pairwise interactions among multiple swimmers. We find multiple pairwise interactions randomize the directions of bacteria and contribute to the active noise in their trajectories. We compare theoretical and numerical prediction to experiments for PA01 swimming at fluid interfaces.

\section{Pair interactions between a tracer and a swimmer}

Chisholm et al. \cite{chisholm_driven_2021} theoretically describe the flow fields that can be generated by an interfacially-trapped, motile bacterium.
This description reveals that, sufficiently far from the bacterium, the flow in the interface is generally dominated by a superposition of two dipolar flow modes, named the  ``\textbf{S}''  and ``\textbf{B}'' modes \cite{chisholm_driven_2021}.
These modes result from a multipole expansion of the velocity field due to a self-propelled object adsorbed to the interface.
The \textbf{S} mode (called the ``\(\bm S^{\parallel}\) mode'' in \cite{chisholm_driven_2021}) is an incompressible interfacial stresslet determined by the balance of flagellar thrust and the counteracting drag on the bactrium's body projected on the interfacial plane. The \textbf{B} mode, on the other hand, is associated with Marangoni stresses generated by the protrusion of the bacterium’s body and flagellum into the surrounding bulk fluid phases.

Superposition of the \textbf{S} and \textbf{B} modes leads to a fore-aft asymmetric flow shown.
Mathematically, each of these modes can be characterized by a magnitude and a direction; they also share a common origin or pole where the fluid velocity is singular.
The magnitude, origin, and direction of the \textbf{S} and \textbf{B} modes were fitted to the observed flow field.
The magnitudes, or strengths, for the \textbf{S} and \textbf{B} modes are $S=0.36\pm$ \SI{0.05}{pN\cdot \micro m} and $B=0.82\pm$\SI{0.02}{pN\cdot \micro m}, respectively.
The direction of the stresslet is offset from the swimming direction by an angle $\phi^S=-8.0\pm2.3^{\circ}$ for the \textbf{S} mode and $\phi^B=192.7\pm4.0^{\circ}$ for the \textbf{B} mode. 
The ``pole'' of the multipole expansion giving the \textbf{S} and \textbf{B} modes is located slightly behind the body’s center at $(0,\delta y)$ with $\delta y=-1.33\pm\SI{0.14}{\micro m}$.
The resulting flow field reflects the forces distributed around the swimmer, which are related to the configuration of the trapped bacteria. Furthermore, the form of this flow reveals that the bacteria swim on an incompressible fluid interface with negligible surface viscosity.

Here, we focus on the implications of bacteria on enhanced transport of tracer particles in the interface using this flow field. The Lagrangian displacement or ``drift'' of a tracer particle in a bacterial flow field (see schematics in Fig.\ref{fig1}e) as a function of time with initial tracer position $\vec{x}_{p,0}$ is given by
\begin{equation}
\Delta\vec{x}_p(t)=\vec{x}_p(t)-\vec{x}_{p,0}=\int_{0}^{t}\vec{u}(\vec{x'}(t'),\vec{q}(t'))dt',
\label{eqn3}
\end{equation}
where $\vec{x'} = \vec x_p(t) - \vec x_s(t)$ is the position of the tracer particle with respect to the hydrodynamic origin of the bacterium (\(r = \norm{\vec x'}\)), $\vec{q}=\langle-\sin{\phi}(t'),\cos{\phi}(t')\rangle$ is a unit vector giving the bacterium’s swimming direction, and $\phi$ is the angle between the bacterium's swimming direction and the $y$ axis. The fluid velocity $\vec{u}(\vec{x'},\vec{q})$ is calculated by superposition of the velocity contributions from the \textbf{S} and \textbf{B} modes, $\vec{u}^S(\vec{x'},\vec{q}^S)$ and $\vec{u}^B(\vec{x'},\vec{q}^B)$, given by
\begin{equation}
    \vec{u}^S(\vec{x}^{\prime},\vec{q}^{S}) = \frac{S}{4\pi\bar{\mu}} \left[
    \frac{3 {(\vec q^S \cdot \vec x')}^2 \vec x'}{r^5} - \frac{(\vec q^S \cdot \vec x') \vec q^S + \vec x'}{r^3}
  \right],
\end{equation}
\label{si:eq4}
and
\begin{equation}
\label{SI:eq5}
\vec{u}^B (\vec{x}^{\prime},\vec{q}^{B})  = \frac{B}{8\pi\bar\mu}\left[\frac{\vec q^B}{r^2} - \frac{2 (\vec q^B \cdot \vec x') \vec x'}{r^4} \right],
\end{equation}
respectively, where
$\bar{\mu}$ is the average viscosity of the bulk fluids. The unit vectors
\[ \vec{q}^S=\langle-\sin{(\phi+\phi^S)},\cos{(\phi+\phi^S)}\rangle \]
and
\[ \vec{q}^B=\langle-\sin{(\phi+\phi^B)},\cos{(\phi+\phi^B)}\rangle \]
define the orientations of the \textbf{S} and \textbf{B} modes relative to the $y$ axis, where $\phi^S$ and $\phi^B$ are the orientation angles of the \textbf{S} mode and the \textbf{B} mode with respect to the swimming orientation observed in experiment.
The tracer displacements can be calculated by,
\begin{equation}
\label{SI:sum}
\Delta \vec{x}_p=\Delta \vec{x}^S_p (\vec{x'}, \vec{q}^S) +\Delta \vec{x}^B_p(\vec{x'}, \vec{q}^B).
\end{equation}

In the limit of large tracer-swimmer separation distance, tracer displacements are small compared to this distance, $r\gg\Delta x_p$, allowing the approximation $\vec{x'} \approx \vec{x}_{p,0} - \vec{x}_s$. This assumption allows analytical integration of eqn (\ref{eqn3}) to find the tracer displacement by the velocity contributions from the \textbf{S} and \textbf{B} modes, $\Delta \vec{x}^S_p$ and $\Delta \vec{x}^B_p$, respectively. We perform this integration to find $\Delta \vec{x}^S_p$ and $\Delta \vec{x}^B_p$ for the dipolar strengths, positions and orientations fitted to Fig.\ref{fig1}b. 
Similarly, we generate an experimental tracer trajectory from integration over the  displacement vectors $\Delta \vec{x}_p$ extracted from the experimental data. Experimental displacement $\Delta \vec{x}_p (t)$ during time interval of $\Delta t$ is approximated as $\vec{u}(\vec{x'}(t),\vec{q}(t))\Delta t$ by properly shifting and rotating the velocity vectors from the measured flow field. The time increment $\Delta t$ is chosen based on the spacing of velocity vectors.

\subsection{Tracer displacements by a bacterium swimming in a straight line}

We first consider the case of a bacterium swimming along a straight line with  $\vec{q}=\vec{\hat{e}}_y$ at constant speed $v=\SI{10}{\micro m\cdot s^{-1}}$. A tracer initially located at $\vec{x}_{p,0}=\langle x_p,0\rangle$ is displaced by the swimmer moving along a straight line from $\vec{x}_{s}=\langle 0,-L/2\rangle$ to $\langle 0,L/2\rangle$. Tracers are advected by the streamlines in the flow generated by the swimmer.

To construct the `experimental' tracer path, displacements of a tracer at short lag time $\Delta t$, approximated as $\Delta \vec{x}_p (t)=\vec{u}(\vec{x'}(t),\hat{\vec{e}}_y)\Delta t$, are obtained by extracting velocity vectors at position $\vec{x'}$ from the measured flow field aligned along $y$ axis. 
These displacement vectors are summed for the swimmer moving along a straight path with $\vec{x}_s(t)=\langle x_{p,0},-y_s(t)\rangle$ over the range $y_s=-L/2$ to $L/2$, where $L=\SI{120}{ \micro m}$. 
This length corresponds to the size of the domain over which the flow field was measured.
The time increment used in this calculation is $\Delta t=L/N_b v$, where $N_b$ is the number of the integrated displacement vectors equal to the number of grid points along the straight paths, $N_b=50$.

Analytical prediction of the tracer displacement relies on evaluating the integrals in eqn \eqref{eqn3} respectively for $\Delta \vec{x}^S$ and $\Delta \vec{x}^B$ and adding the results to represent the tracer path. The $y$ location of swimmer changes over time as $y_s(t)=y_i+vt$. 
The initial position of the swimmer with respect to the tracer $y_i$ is defined by the $y$-positions of \textbf{S} and \textbf{B} modes with respect to the center of the cell body, thus $y_i=-L/2+\delta y$. Assuming \textbf{S} and \textbf{B} modes are aligned with $y$ axis ($\phi^S=0$ and $\phi^B=\pi$), a closed form of $\vec{x}_p(t)$ can be obtained, with the contribution from the \textbf{S} mode given by,
\begin{multline}
  \label{eq:si7}
   \vec{x}_p^S(t) =  \vec{x}_{p,0}  - \frac{S}{4\pi\bar{\mu} v} \left[
    \frac{-x_{p,0}y_s}{(x_{p,0}^2+y_s^2)^{3/2}} {\hat{\vec{e}}}_x\right. \\
    + \left.\frac{y_s^2}{(x_{p,0}^2+y_s^2)^{3/2}} {\hat{\vec{e}}}_y
  \right]^{y_s=y_i+vt}_{y_s=y_i}.
\end{multline}
Similarly, the contribution from the \textbf{B} mode is given by
\begin{equation}
  \label{eq:si8}
   \vec{x}_p^B(t) = \vec{x}_{p,0} + \frac{B}{8\pi\bar{\mu} v} \left[
    \frac{x_{p,0}}{x_{p,0}^2+y_s^2} {\hat{\vec{e}}}_x -
    \frac{y_s}{x_{p,0}^2+y_s^2} {\hat{\vec{e}}}_y
  \right]^{y_s=y_i+vt}_{y_s=y_i}.
\end{equation}

\begin{figure}[t]
\centering
\includegraphics[width=.70\linewidth]{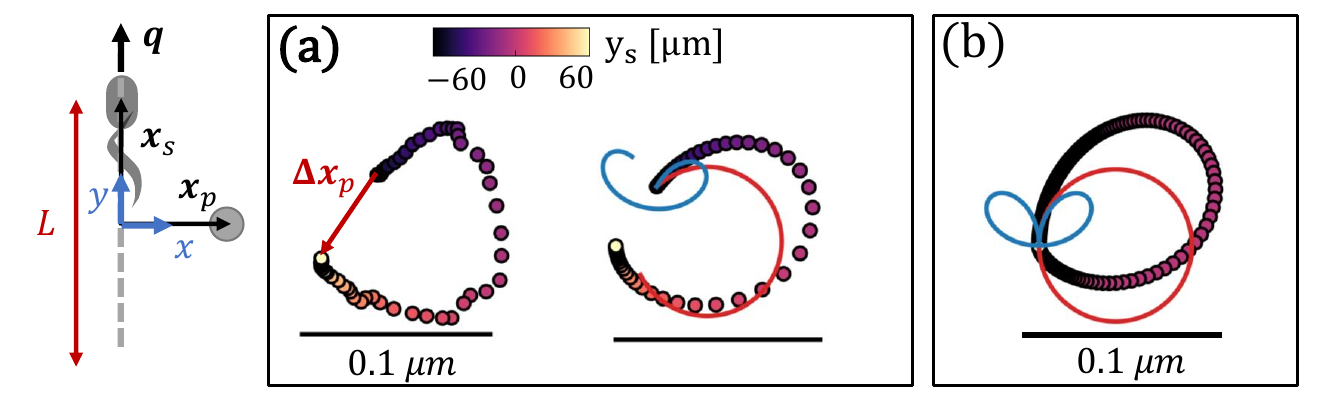}
\caption[Hydrodynamic interactions with a straight swimmer]{\textbf{Hydrodynamic interactions with a straight swimmer.} Schematic: A tracer with initial position $\vec{x}_{p,0}=\langle\SI{20}{ \micro m},0\rangle$ (grey circle) interacts with a pusher bacterium moving along a straight trajectory $x_s=\langle 0,y_s(t)\rangle$.
(a) Tracer path when interacting with a bacterium swimming from $\SI{-60}{\micro m}\leq y_s\leq\SI{60}{\micro m}$. 
Left: Tracer path constructed from experimental flow field. Right: Corresponding prediction by superposition of displacements by \textbf{S} (blue curve) and \textbf{B} (red curve) modes. Circle colors indicate the bacteria's position (heat map). (b) Tracer path forms a closed loop for an infinite bacterial trajectory. 
The colors indicate corresponding bacteria’s position (heat map) for bacteria moving from $y_s=-10^3 \si{\micro m}$ to $y_s=10^3\si{\micro m}$.}
\label{fig3}
\end{figure}

As suggested by the form of the flow (Fig. \ref{fig1}b), when the swimmer approaches and passes the tracer, the tracer is pushed away by the outflow in front of the swimmer and subsequently is pulled toward the swimmer by the inflow at the rear. This yields a loopy tracer path as shown in Fig. \ref{fig3}a. 
The contributions from each mode are also calculated from numerical integration of both modes, with the multiple lobes in the \textbf{S} mode generating a two-lobed "lima-bean" shaped path for the tracer (blue curve), while the \textbf{B} mode generates a symmetric loop with a net displacement in the direction opposite the bacterial swimming direction (red curve). These displacements superpose to yield an asymmetric tracer path with a tilted loop, with an opening $\Delta \vec{x}_p$ due to the finite swimmer path. 
As expected, for this straight swimmer, the tracer path closes as the bacterial trajectory elongates, generating zero net displacement for an infinite swimmer trajectory (Fig. \ref{fig3}b). 
The closed loop is also observed for swimmers in bulk fluids with infinite trajectory \cite{pushkin_fluid_2013}.
Theoretically, the shape and size of the tracer's path depends on swimmer’s speed, relative strength of dipolar modes, path length, and initial position of the tracer.

\subsection{Tracer displacements by a bacterium swimming in a circular path}
A related approach allows prediction of tracer displacements near a pusher swimming in circles, $\vec{x}_s=R_s \cos(\Omega t)\hat{\vec{e}}_x+R_s \sin(\Omega t)\hat{\vec{e}}_y$, where $R_s$ is the radius of the swimmer’s circular trajectory and $\Omega$ is its rotational velocity. For tracers far from the bacterium, and for swimmers that rotate at nearly fixed positions with highly curved trajectories ($r\gg R_s$), the position vector characterizing tracer-swimmer separation, $\vec{x'}$ can be assumed to be independent of time ($\vec{x'}=\vec{x}_{p,0}$), while the swimmer’s orientation changes periodically as $\vec{q}=-\sin(\phi)\hat{\vec{e}}_x+\cos(\phi)\hat{\vec{e}}_y$, where $\phi$ is the rotation angle of the swimmer, $\phi(t)=\Omega t$. 

The experimental tracer path can be graphed by integrating over displacement vectors $\Delta \vec{x}_p(t)=\vec{u}(\vec{x'},\phi(t))\Delta t$ from the measured flow field at a fixed position $\vec{x'}$ with the flow field rotating CCW by $\Delta \phi$ at each time interval from $\phi=0$ to $\phi=2\pi$.
To achieve this, the measured flow field is reconstructed on a polar grid of equivalent radial spacing and equivalent angle intervals of $\Delta \phi=\pi/15$ (details see S1, ESI). 
Equivalent to integrating over velocity vectors at fixed $\vec{x'}$ from a rotating field, we extract the velocity vectors in a circular path from a flow field oriented in the $y$ direction. The velocity vectors at $\vec{x'} = \langle x_{p,0}, \phi (t)\rangle$ in polar coordinate are rotated CCW by $\phi (t)$ and integrated from $\phi = 0$ to $2\pi$. 
The spacing between the velocity vectors along the circular path determines the  time interval over which displacement is measured as $\Delta t=\Delta \phi/\Omega$, where $\Omega$ is approximated as $2\ s^{-1}$.

Similarly, tracer displacements can be constructed analytically. The relative distance between the tracer and the two dipolar modes is assumed to remain fixed, $\vec{x'}=\langle x_{p,0},0\rangle$, while the bacterium’s rotational angle changes as $\phi=\phi_i+\Omega t$, where $\phi_i$ is the initial orientation of the bacterium with respect to the $y$ axis; $\phi_i$ is corrected to $\phi_i^S=\phi^S$ for the \textbf{S} mode and $\phi_i^B=\phi^B$ for the \textbf{B} mode to account for the differing alignment of the \textbf{S} and \textbf{B} modes. 
Integration of $\vec{u}^S$ over circular paths of the swimmer in the limit of $r\gg R_s$ yields,
\begin{equation}
  \label{eq:si9S}
  \vec{x}_p^S(t) = -\frac{S}{16\pi\Omega\bar{\mu}} \frac{1}{x_{p,0}^2} \left[
    2\sin{2\phi} {\hat{\vec{e}}}_x + \cos{2\phi} {\hat{\vec{e}}}_y
  \right]^{\phi=\phi_i^S+\Omega t}_{\phi=\phi_i^S} + \vec{x}_{p,0},
\end{equation}
while a similar integration of $\vec{u}^B$ yields 
\begin{equation}
  \label{eq:us_si10B}
  \vec{x}_p^B(t) = \frac{B}{8\pi\Omega\bar{\mu}} \frac{1}{x_{p,0}^2} \left[
    -\cos{\phi} {\hat{\vec{e}}}_x + \sin{\phi} {\hat{\vec{e}}}_y
  \right]^{\phi=\phi_i^B+\Omega t}_{\phi=\phi_i^B} + \vec{x}_{p,0}.
\end{equation}

\begin{figure*}[h]%
\centering
\includegraphics[width=.92\linewidth]{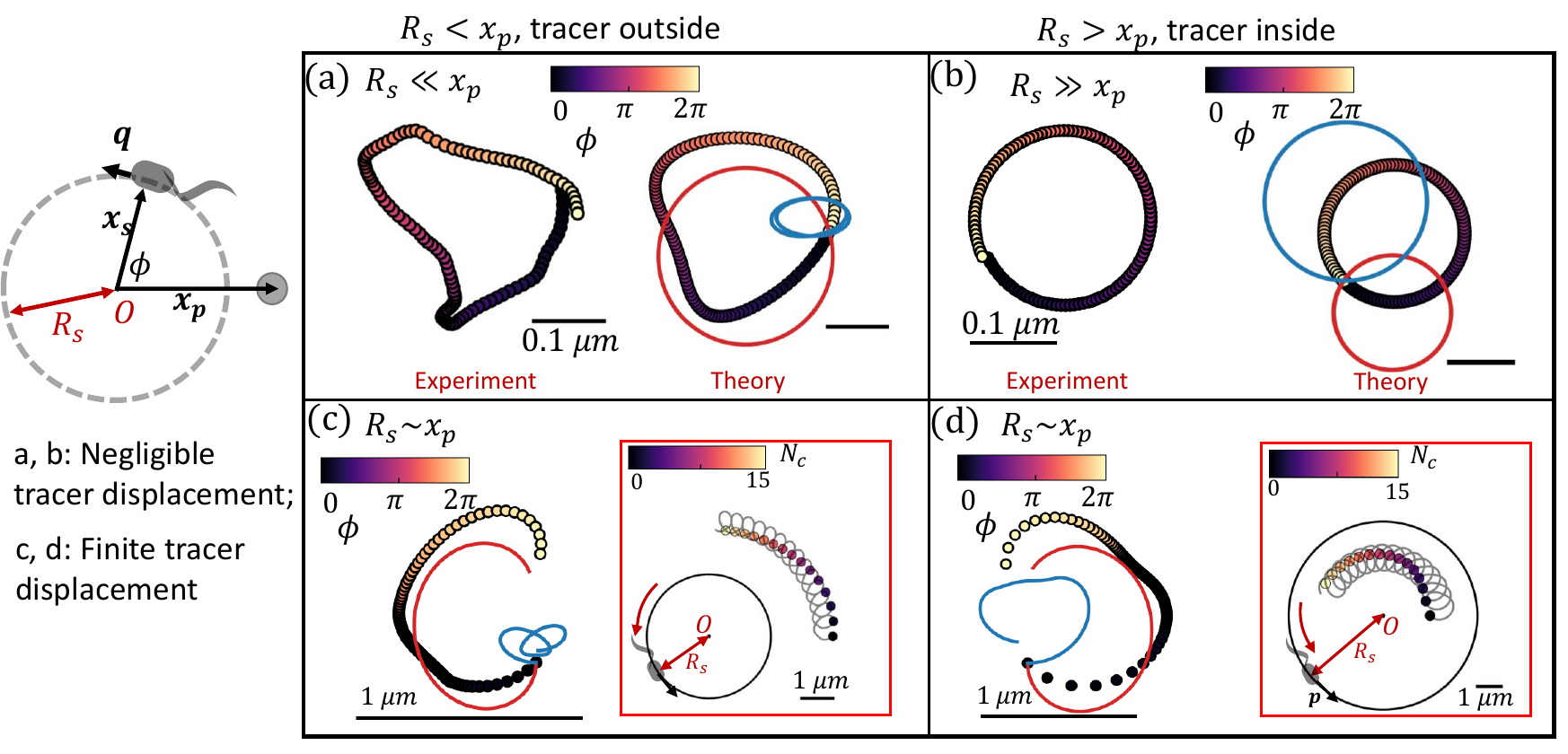}
\caption[Hydrodynamic interactions with a circular swimmer.]{\textbf{Hydrodynamic interactions with a circular swimmer.} Schematic: Tracer positioned at $\vec{x}_{p,0}$ (grey circle) interacting with a bacterium swimming along a circular trajectory of radius $R_s$. (a) Tracer far from the origin of the swimmer’s circle. Left: Closed loop tracer path constructed from experimental displacements after bacterium completes one rotation for $\vec{x}_{p,0}=\langle \SI{15}{\micro\meter},0\rangle$ and $R_s\ll x_p$. Right: Corresponding tracer path predicted by superposition of \textbf{S} (blue curve) and \textbf{B} modes (red curve). Circle colors indicate swimmer’s rotation angle (heat map). (b) Tracer at origin of the swimmer circular path. Left: Closed circular tracer path constructed from experimental displacement field for a bacterium performing one rotation for $R_s=\SI{15}{\micro m}$ and $\vec{x}_{p,0}=\langle0,0\rangle$. Right: Corresponding prediction from superposition of \textbf{S} (blue curve) and \textbf{B} modes (red curve). Circles colors: same as in (b). (c,d) Finite particle displacements significant for $R_s\sim x_p$ generate net tracer displacement for circular swimmers after one bacteria rotation. (c) Predicted tracer displacement for particle initially at $\vec{x}_{p,0}=\langle\SI{4}{\micro\meter},0\rangle$ for $R_s=\SI{2}{\micro \meter}$, outside of the swimmer’s circle from superposition of \textbf{S} (blue) and \textbf{B} (red) modes. Circle colors: same as in (a). (d) Predicted tracer displacement for particle initially at $\vec{x}_{p,0}=\langle \SI{4}{\micro\meter},0\rangle$ for $R_s=\SI{6}{\micro m}$, inside the swimmer’s circle, from superpositions of \textbf{S} (blue) and \textbf{B} (red) modes. Circle colors: same as in (b). Insets to (c) and (d): Persistent swimming on circular paths generates tracer displacement on helical paths. Circle colors indicate rotation number $N_c$ (heat map). }
\label{fig4}
\end{figure*}

As the swimmer completes a circle, the tracer moves in a closed loop, as can be predicted from analysis in eqn (\ref{eq:si9S}) and (\ref{eq:us_si10B}). 
The \textbf{S} mode leads to two ellipse-shaped tracer displacement loops for each period, (Fig. \ref{fig4}a, blue curve), while the \textbf{B} mode generates a single circular loop (Fig. \ref{fig4}a, red curve). Because of this period doubling, the \textbf{S} mode generates smaller tracer displacements than does the \textbf{B} mode for interaction times that do not correspond to multiples of the bacteria’s period of rotation. 
Superposition of the displacements generated by these two modes results in a closed loop with a rounded triangular shape (Fig. \ref{fig4}a, right); the size of this loop scales as $\Omega^{-1}r^{-2}$.
The predicted tracer path agrees with that extracted from the measured flow field (Fig. \ref{fig4}a, left). 
The tilted tip of the triangle results from the differing orientations of the \textbf{S} and \textbf{B} modes ($\phi^B$ and $\phi^S$).

We also consider the tracer placed at the center of circular path of bacteria with $x_{p,0}\ll R_s$, thus the distance between the swimmer and the tracer remains unchanged ($r= R_s$) over the time, while the direction of their relative separation $\vec{x'}$ remains perpendicular to the swimmer orientation $\vec{q}$. 
Thus, the tracer follows a circular path with a constant speed determined by its initial separation from swimmer ($\vec{x'}_i=\langle -R_s, 0\rangle$) and swimmer's initial orientation ($\vec{q}_i=\hat{\vec{e}}_y$). 
The velocity vector at $\vec{x'}_i$ is extracted from the measured flow field, rotated continuously in a CCW sense by an interval of $\Delta \phi=\pi/12$ and integrated from $\phi=0$ to $2\pi$. 
The nearly-circular resulting tracer path extracted from the experimental flow field is shown in Fig. \ref{fig4}b left. Analytical integration of the velocity $\vec{u}^S$ yields a circular path of tracer given by, 
\begin{equation}
  \label{eq:si11}
  \vec{x}_p^S(t) = \frac{S}{4\pi\Omega\bar{\mu}} \frac{1}{R_s^2} \left[
    \sin{\phi} {\hat{\vec{e}}}_x - \cos{\phi} {\hat{\vec{e}}}_y
  \right]^{\phi=\phi_i^S+\Omega t}_{\phi=\phi_i^S} + \vec{x}_{p,0}.
\end{equation}
Similarly, integration of $\vec{u}^B$ yields,
\begin{equation}
  \label{eq:us_si12}
  \vec{x}_p^B(t) = \frac{B}{8\pi\Omega\bar{\mu}} \frac{1}{R_s^2} \left[
    \cos{\phi} {\hat{\vec{e}}}_x + \sin{\phi} {\hat{\vec{e}}}_y
  \right]^{\phi=\phi_i^B+\Omega t}_{\phi=\phi_i^B} + \vec{x}_{p,0}.
\end{equation}
The tracer displacements from each mode and their superposition are plotted in Fig. \ref{fig4}b right. 
Unlike when the tracer is located far away from the bacterium, in this case, the \textbf{S} mode generates a single circular loop in phase with path generated by \textbf{B} mode, yielding a circular tracer path with a rotational period equal to that of the swimmer. 
In both limits, as the swimmer completes one rotation, the tracer is displaced along a closed loop with no net displacement. The tracer moves in an oscillatory manner, with limited impact on interfacial transport.

\subsection{Tracer displaced by a bacterium in close proximity}
While no net hydrodynamic displacement of tracers is induced by a swimmer moving in complete circles when swimmer and tracer are well-separated, a finite displacement is predicted for tracers and swimmers in closer proximity. 
In this limit, changes in the separation distance between the swimmer and the tracer, $\vec{x'}$, cannot be neglected when calculating the fluid velocity at the tracer’s location. In this regime, the determination of the displacement of a tracer by a swimmer requires numerical integration. We perform this integration for a bacterium swimming in CCW circles with its orientation $\vec{q}$ normal to its position vector $\vec{x}_s$.
The induced velocity $\vec{u}(\vec{x'}(t), \vec{q}(t))$ is integrated to find the position of the tracer and to calculate $\vec{x'}(t)$. 
Thus, we numerically solve eqn \eqref{eqn3} using an Euler integration method with a step size of $\Delta t=10^{-4} s$ to calculate the changes in their separation distance, $\vec{x'}(t)=\vec{x}_p(t)-\vec{x}_s(t)$, due to the finite displacements of tracers. With the step size of $\Delta t=10^{-4} s$, the error of the numerical integration of the tracer displacement at each rotational circle is negligible compared to tracer displacement. 

We find that the changes in $\vec{x'}(t)$ break the symmetry of the system and generate a directed tracer displacement after full rotation of the swimmer. A tracer originally positioned just outside of the swimmer’s circle moves along a loop-shaped path depicted in Fig. \ref{fig4}c with an important feature; this loop has a finite opening. We find similar results for a tracer positioned inside of the swimmer’s circle (Fig. \ref{fig4}d). 
Furthermore, continuous swimming over many circles yields a looped tracer trajectory which orbits around the swimmer over long times, as is shown in the insets to Fig. \ref{fig4}c and d, where each point represents the position of the tracer at the beginning of each period of the swimmer’s circular motion. 
However, no attraction or repulsion is detected over a period of swimmer motion due to the divergence free nature of the incompressible interface. 
This looped motion generates significant net displacements at long lag times, providing a mechanism for directed tracer motion and enhanced tracer dispersion.

\section{Pair interactions between swimmers}
We now consider cases of two bacteria whose circular paths are perturbed via advection and reorientation by each other’s flow fields. The position vector of a bacterium swimming in the laboratory frame with initial position $\vec{x}_{s,0}$ is given by 
\begin{equation}
  \label{eqn4}
  \vec{x}_s(t) = \int_{0}^{t}(v\vec{q}+\vec{u}'(\vec{x}_s-\vec{x}'_s,\vec{q}'))dt'+\vec{x}_{s,0},
\end{equation}
where $\vec{u}'(\vec{x}_s-\vec{x}'_s,\vec{q}')$ denotes the fluid velocity generated by a neighboring swimmer with orientation $\vec{q}'(t)$ and position $\vec{x}_s'(t)$.
The contributions to torque associated with the bacterium’s interfacially-trapped state and flagellar rotation drive a change in orientation ${d\vec{q}_{int}}/{dt}=\Omega\hat{\vec{e}}_{\theta}$ in polar coordinates. In addition, the vorticity $\vec{\omega}'$ and strain rate $\vec{E}'$ at position $\vec{x}_s -\vec{x}_s'$ generated by the neighbor changes the swimmer’s orientation. Assuming an ellipsoidally shaped bacterium, this re-orientation is given by,
\begin{equation}
  \label{eqn5}
 \frac{d\vec{q}_{ext}}{dt}= \frac{1}{2}\vec{\omega'}\times\vec{q}+ \Gamma^* \vec{q} \cdot\vec{E'}\cdot(\vec{I}-\vec{q}\vec{q}),
\end{equation}
where $\vec{I}$ is the identity matrix, $\Gamma^{*}=[(\gamma^*)^2-1]/[(\gamma^*)^2+1]$ is the geometrical factor of bacterium, and $\gamma^*$ is the total aspect ratio, including the cell body and the flagellum \cite{pedley_hydrodynamic_1992, drescher_fluid_2011}. 
Flagellated bacteria typically have $\gamma^*\gg1$, thus $\Gamma^*\sim 1$.
The instantaneous change in each bacterium’s swimming direction can thus be predicted by the sum
\begin{equation}
  \label{eqn_rot}
\frac{d\vec{q}}{dt}=\frac{d\vec{q}_{int}}{dt}+\frac{d\vec{q}_{ext}}{dt}.
\end{equation}
Notably, the extrinsic reorientation due to hydrodynamic interaction decays as $|\vec{x}_s-\vec{x}'_s|^{-3}$, and it is only comparable to the rate of intrinsic rotations 
($1-\SI{10}{s^{-1}}$) for swimmers separated by less than ten microns. 
We study two swimmers at such separation distances. In this regime, the reorientating effect of a neighboring swimmer $\sim v \Delta t^2 \, {d\vec{q}_{ext}}/{dt}$ is more pronounced than its advection effect $\sim\vec{u}'\Delta t$ during an interaction time $\Delta t>\SI{0.1}{s}$ via hydrodynamic interaction.

\begin{figure}[h]
\centering
\includegraphics[width=.5\linewidth]{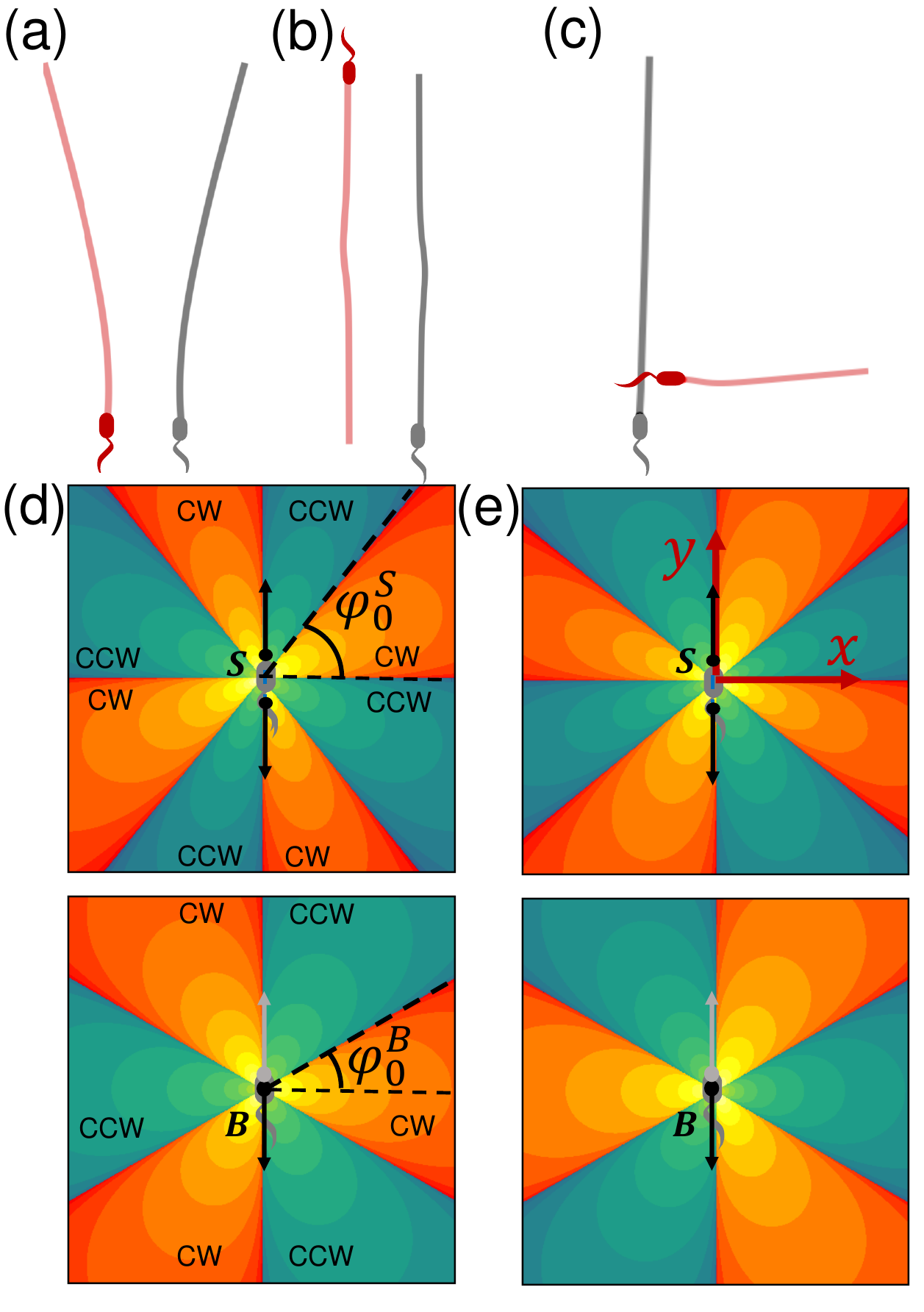}
\caption[Reorientation of neighboring swimmer due to the
\textbf{S} and \textbf{B} modes of a source swimmer]{\textbf{Reorientation of neighboring swimmer due to the
\textbf{S} and \textbf{B} modes of a source swimmer}. (a) Trajectories of pairs of initially parallel swimmers
initially oriented along the $y$ axis. (b) Trajectories of pairs of
initially anti-parallel swimmers initially swimming in \(\pm y\)
directions. (c) Trajectories of pairs of initially perpendicular
swimmers. (d) Reorientation of a neighboring swimmer moving parallel or
anti-parallel to a source swimmer. The source swimmer moves in the $+y$
direction at the origin. Upper panel: rotational velocity field
\(\Omega^{S}\). Lower panel: rotational velocity field \(\Omega^{B}\).
Red zones: neighbor rotates in the CW sense. Blue zones: neighbor
rotates in the CCW sense. The heat map indicates the magnitude of the
angular velocity with brighter color indicating larger angular velocity. (e) Reorientation of a neighboring swimmer moving along $+x$ direction,
perpendicular to a source swimmer that moves in the $+y$ direction at the
origin. Panels and colors similar to (d).}
\label{reorient}
\end{figure}

\subsection{Reorientation of swimmers with initially straight swimming paths}
When two bacteria approach, each is reoriented with a rotational
velocity determined by the vorticity and strain rate generated by its
neighbor, described in eqn (\ref{eqn5}). The net effect of such
rotation on swimmers' long-term trajectories depends on the swimmers'
initial relative positions and orientations and the shapes of their
swimming paths. We focus here on swimmers
moving along initially straight paths. Parallel swimmers positioned initially adjacent to each other and
moving at the same speed are repelled from each other, Fig. \ref{reorient}a. In this case, the interaction is persistent, and the net effects on the swimmers'
long-term paths are pronounced. On the other hand, anti-parallel
swimmers deviate only weakly from their straight paths. As they
approach, they rotate away from each other. Their sense of rotation
reverses as they pass each other. They rotate back to co-alignment as
they move away from each other, eventually resuming their straight
paths, Fig. \ref{reorient}b. In this case, the interaction is transient, and the net effects
on swimming paths are weak. To understand swimmers on perpendicular paths, Fig. \ref{reorient}c, we must first discuss the reorientation of swimmers more generally.

To illustrate the predicted range of behaviors in greater detail, we discuss the reorientation for swimmers moving initially on
straight paths as a function of initial position. We define a source
swimmer that moves along the $y$ axis and reorients its neighbor. The
neighbor can be rotated counterclockwise (CCW) or clockwise (CW)
depending on its position and orientation with respect to the source.
This sense of rotation changes
with the polar angle \(\varphi\ \) of the neighbor in the $x$-$y$ plane in Fig. \ref{reorient}. We summarize
the rotational velocity fields \(\Omega^{S}\) and \(\Omega^{B}\) for the
\textbf{S} and \textbf{B} modes, respectively. For each mode, we depict
zones in which a neighbor would undergo CW rotation in red and CCW
rotation in green. Rotation generated by the \textbf{S} mode is
determined by the vorticity field and the strain rate of the source
swimmer, whereas, the \textbf{B} mode, being irrotational, reorients its
neighbors only via its local strain rate.

\emph{Swimmers that move initially in the same direction:} In Fig.\ref{reorient}a
we consider adjacent co-aligned swimmers, (for which  \(\varphi\ \)$=0$), moving in the \(y\) direction. In Fig.\ref{reorient}d,
we report
\begin{equation}
  \label{eqn_rorS}
 \Omega^{S} = \frac{S}{4\pi\bar\mu}\left(\frac{9xy}{r^{5}} - \frac{15xy^3}{r^{7}}\right),
\end{equation}
where \(\Omega^{S} > 0\) indicates clockwise (CW) rotation of the
neighboring bacteria due to \textbf{S} mode of the source swimmer at the
origin. Consider a neighboring swimmer located in the quadrant
\(0 < \varphi\  < \pi/2\). For \(0 < \varphi < \varphi_{0}^{S}\)
the neighbor rotates CW, whereas for
\(\varphi_{0}^{S} < \varphi < \pi/2\), the neighbor's sense of
rotation changes to counterclockwise (CCW). For
\(\varphi = 0\), \(\varphi = \varphi_{0}^{S}\), and \(\varphi = \pi/2\), the
angular velocity passes through zero, and the neighbor does not rotate.
In this quadrant, analysis reveals that the polar angle at which a neighbor experiences no rotation by the \textbf{S} mode  \(\varphi_{0}^{S} = {50.76}^{\circ}\). The
angular velocity \(\Omega^{S}\ \)is anti-symmetric with respect to the \(x\)
and \(y\) axes. These rotations can cause swimmers initially moving on
parallel paths to attract or repel. In the quadrant
\(0 < \varphi < \pi/2\), CW rotation generates repulsion and
CCW rotation generates attraction. However, for neighbors in the quadrant
\(\pi/2 < \varphi < \pi\), CCW rotation generates repulsion and
CW rotation generates attraction. 
Angular velocity maps generated by the
\textbf{B} mode of the source swimmer are graphed for co-aligned
neighbors in the bottom panel, given by
\begin{equation}
  \label{eqn_rorB}
\Omega^{B} = \frac{B}{4\pi\bar\mu}\left( \frac{x^{3} - 3xy^{2}}{r^{6}} \right). 
\end{equation}
Considering again the quadrant \(0 < \varphi < \pi/2\),
neighbors rotate in the CW sense and repel for
\(0 < \varphi < \varphi_{0}^{B}\), whereas neighbors rotate CCW and
attract for \(\varphi_{0}^{B} < \varphi < \pi/2\); analysis shows
\(\Omega^{B}\) passes through zero for
\(\varphi = \varphi_{0}^{B} = 30^{\circ}\). The rotational velocity
\(\Omega^{B}\ \)is symmetric with respect to the $x$ axis, anti-symmetric
with respect to the $y$ axis, and has no root at \(\varphi = 0\), but
rather favors CW rotation that generates the repulsion.

\emph{Swimmers that move initially in opposite directions:} Neighbors
that swim anti-parallel to the source at the origin experience identical
angular velocities as parallel swimmers for both the \textbf{S} and
\textbf{B} modes (Fig.\ref{reorient}b). However, for this direction of swimming,
zones of attraction and repulsion are reversed, i.e. CW rotation
generates attraction, and CCW rotation generates repulsion for
\(0 < \varphi < \pi/2\).

\emph{Swimmers with initially perpendicular arrangement:} We can now discuss the case of a source
swimmer initially orientated along $y$ axis interacting with a neighboring
swimmer initially swimming parallel to the $x$ axis (Fig.\ref{reorient}c). Neighbors
swimming in the positive or negative $x$ directions experience the same rotational
velocity fields, as depicted in Fig.\ref{reorient}e. For this orientation, the
angular velocity generated by the \textbf{S} mode of the source swimmer
is
\begin{equation}
  \label{eqn_rorBB}
\Omega^{S} = - \frac{S}{4\pi\bar\mu}\left( \frac{6xy}{r^{5}}-\frac{15xy^3}{r^{7}}  \right).
\end{equation}
For \(0 < \varphi < \varphi_{0}^{S}\) the neighbor rotates CCW, whereas
for \(\varphi_{0}^{S} < \varphi < \pi/2\), the sense of rotation
changes to CW, with \(\varphi_{0}^{S} = \SI{39.24}{\degree}\). The angular
velocity for the \textbf{B} mode is given by
\begin{equation}
  \label{eqn_rorSS}
\Omega^{B} = - \frac{B}{4\pi\bar\mu}\left( \frac{x^{3} - 3xy^{2}}{r^{6}} \right), 
\end{equation}
which causes the neighbor swimmer to rotate CCW for
\(0 < \varphi < \varphi_{0}^{B}\), and rotate CW for
\(\varphi_{0}^{B} < \varphi < \pi/2\)
(\(\varphi_{0}^{B} = \SI{30}{\degree}\)).

In summary, when swimmers at interfaces interact, the net effect of their velocity fields depends on the balance of the \textbf{S} and \textbf{B} modes, which is determined by their trapped configuration at the interface, and the instantaneous separation and orientations of the swimmers.

\subsection{Pair interactions between swimmers with circular paths} 
\begin{figure}[t]
\centering
\includegraphics[width=.8\linewidth]{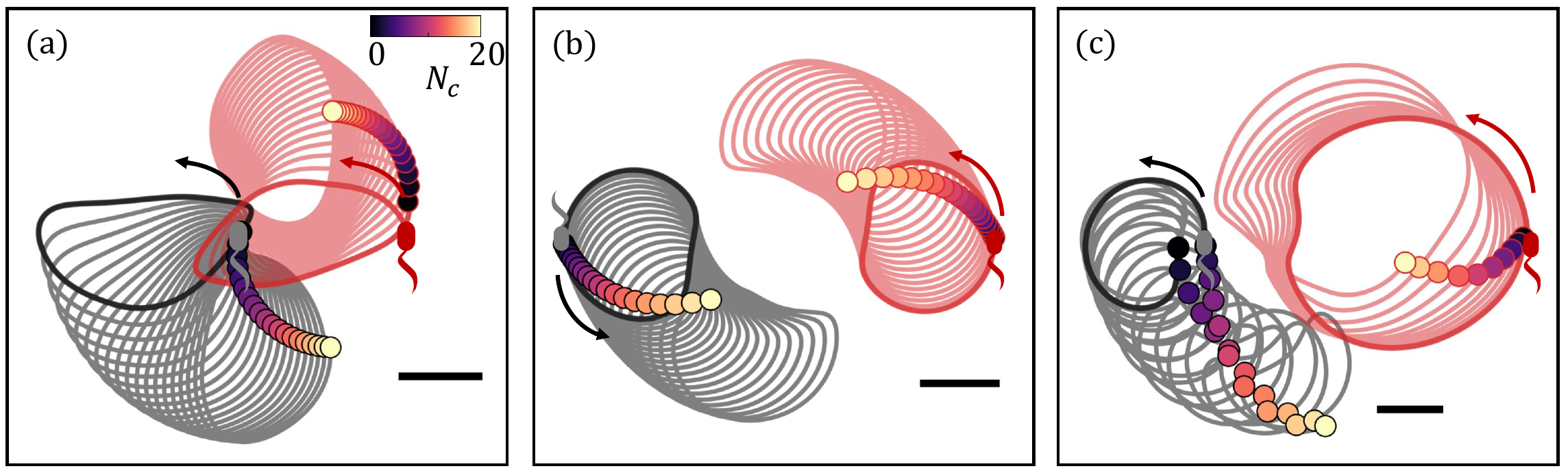}
\caption[Hydrodynamic interactions between swimmer pairs with circular paths]{\textbf{ Hydrodynamic interactions between swimmer pairs with circular paths.} (a) and (b) Hydrodynamic interactions generate helical trajectories; these examples illustrate repulsive net displacements for initially co-oriented swimmers moving along circles with the same radii, and attractive interactions for swimmers with opposing initial orientations. (c) Finite displacements occur for pairs of swimmers with circular trajectories with differing radii, as shown for $R_s$ of \SI{1}{\micro m} and \SI{2}{\micro m}. Scale bars indicate \SI{1}{\micro m} }
\label{fig5}
\end{figure}

Predicted trajectories can deviate significantly from the
swimmer's trajectories in isolation, suggesting that swimmer
hydrodynamic interactions can be an important source of active noise.
For swimmers moving periodically along circular paths, such changes in
orientation and position can result in a small net displacement in each
period. Given that swimmers can interact over many periods, these
effects can result in significant displacements over prolonged
interaction times (Fig. \ref{fig5}). The swimmers can attract or repel each other depending on their initial orientations and positions. This prediction is particularly important, since interfacially trapped pushers typically swim along circular paths.

\section{Tracers interacting with multiple swimmers}
Tracer displacements are known to be significantly enhanced by interactions with multiple swimmers \citep{pushkin_fluid_2013, burkholder_tracer_2017,kurtuldu_enhancement_2011,kasyap_hydrodynamic_2014,kanazawa_loopy_2020}. While such micro-mixing is well studied for swimmers moving along straight paths, the effects of key features of interfacial swimmers remain unprobed, including their curvilinear paths and the eventual decorrelation of their swimming direction by active perturbations \citep{deng_motile_2020}. 
Active perturbations generate diffusive displacements at long lag times and can be described as an active diffusivity $D_{act}$ \citep{deng_motile_2020}. This diffusivity randomly perturbs the tracer-swimmer separation distance and the swimmers’ orientation, and therefore alters tracer advection. While thermal diffusion of the tracer and bacterium can also generate randomizing effects, here we focus on  the role of $D_{act}$, bacterial density, and hydrodynamic interactions.

\subsection{Simulation of tracer displacements}
\begin{figure*}
\centering
\includegraphics[width=0.92\linewidth]{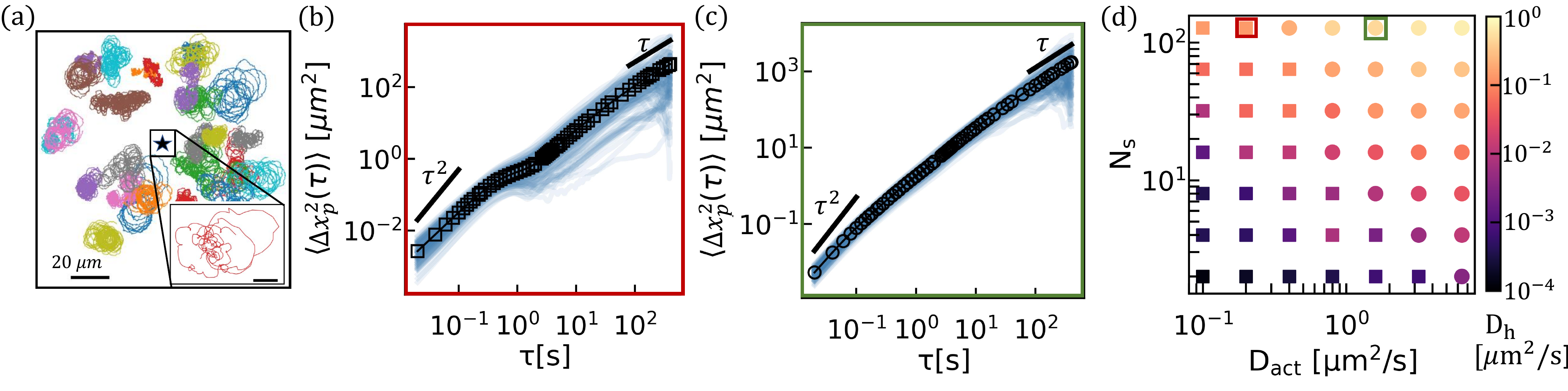}
\caption[Hydrodynamic interactions generate tracer caging/uncaging and enhanced diffusion of tracer particles.]{\textbf{Hydrodynamic interactions generate tracer caging/uncaging and enhanced diffusion of tracer particles.} (a) Schematic of the simulation of a single tracer advected by $N_s=32$ swimmers with active noise in the swimmers’ trajectories within a square domain of side length of $b=\SI{100}{\micro m}$ ($\rho_s=Ns/b^2$). Inset: tracer path. Scale bar is \SI{0.5}{\micro m} (b) Ensemble average of 100 realizations (light blue curves) of MSD of an isolated tracer particle interacting with $N_s=128$ swimmers with active diffusivity $D_{act} =\SI{0.2}{\micro m^2 s^{-1}}$ 
(c) Ensemble average (circles) of 100 realizations (light blue curves) of MSD of an isolated tracer particle interacting with $N_s=128$ swimmers with active diffusivity $D_{act} =\SI{1.6}{\micro m^2 s^{-1}}$.
(d) State diagram summarizing caged states (squares) and uncaged states (circles) for combinations of $N_s$ and $D_{act}$. Symbol color indicates the magnitude of the tracer’s hydrodynamic diffusivity $D_h$. The symbols corresponding to conditions in (b) and (c) are circumscribed by red and green squares.
}
\label{fig6}
\end{figure*}

Here, we simulate displacements of tracer beads moving via pair hydrodynamic interactions with swimmers that follow circular trajectories. In each simulation, a single tracer is placed in the center of an unbounded plane. To this plane, $N_s$ swimming bacteria with \textbf{S} and \textbf{B} modes with dipolar strength of measured values are introduced simultaneously at random locations 
within a square domain of side length $b=\SI{100}{\micro m}$ centered around the tracer's initial location (Fig. \ref{fig6}a). 
We neglect collision and near-field effects between the tracer and the swimmers and set the minimum distance between bacteria and tracer to be $R_{min}=\SI{2}{\micro m}$. The tracer particle displacement generated by $N_s$ swimmers at each time interval of $\Delta t=\SI{0.02}{s}$ is advanced in an explicit Euler scheme. Each swimmer moves with a velocity $v_k$ selected from random numbers which follow a Poisson distribution with a mean value of $\SI{10}{\micro\meter\per\second}$ and with a fixed angular speed $\Omega=\SI{2}{\second^{-1}}$. The radius of curvature for each bacteria $R_c=v_k/\Omega$ has a mean value of \SI{5}{\micro m}. The center of its rotation diffuses with active diffusivity $D_{act}$.
The active noise $\varepsilon(t)$ added to the displacements of swimmers has a decorrelation timescale $\tau=\SI{0.2}{s}$ and active diffusivity $D_{act}$ to preserve the ballistic characteristic of the swimmers' motions at short lag times and diffusive behavior at long lag times.
The interface is dilute, with number density $\rho_s=N_s/b^2<\SI{0.013}{\micro m^{-2}}$, so interactions between swimmers and multi-body interactions are neglected. 
This simulation continues for a time period of $T$ of $\SI{2000}{s}$. For each combination of $N_s$ and $D_{act}$, we repeat this simulation 100 times to provide representative statistics to calculate key quantities. For a bacterium swimming in a cycle of radius $R_c$ with its direction decorrelated in a time scale of $\tau_c$, the maximum active diffusivity was estimated as $D_{act,m}={R_c^2}/{\tau_c} \approx \SI{6}{ \micro m^2 s^{-1}}$ to maintain the circular signature of bacterial path \cite{deng_motile_2020}. 

\subsection{Enhanced diffusion of tracers}
In Fig. \ref{fig4}c and Fig. \ref{fig4}d, we show that a tracer interacting with a bacterium swimming on a circular path moves in a looped trajectory. 
The importance of such tracer-swimmer interactions on micro-mixing can be assessed by measuring the MSD of a single tracer interacting with a swimmer moving on a circular path perturbed by active noise. 
To quantify the effect of this interaction, we study the magnitude of the MSD and its local logarithmic slope
\begin{equation}
  \label{eqn_alpha}
\alpha(\tau)= \frac{\partial \ln\langle\Delta x_p^2 (\tau)\rangle}{\partial\ln \tau}. 
\end{equation}
The tracer MSD shows several similarities with that of the interfacial curly bacteria \citep{deng_motile_2020}. 
For short lag times, tracer and swimmer motion are highly correlated, and the tracer moves ballistically along streamlines in the flow with MSD growing quadratically with lag time. 
At intermediate lag times, \(\alpha<1\) is small indicating that the tracer is “caged” due to the loop-shaped tracer displacement trajectories. After the tracer is released from this dynamic cage, it continues to move superdiffusively with anomalous diffusion exponent $1<\alpha<1.5$. 
Due to the combined effects of active diffusion and convection the tracer displacement eventually becomes diffusive at long lag time, \(\alpha\approx 1\). 
The superdiffusive MSD after cage breaking reveals the directed motion of the tracer at long lag times owing to interactions with the bacteria continuously swimming in a CCW sense at finite separation distance (inset to Fig. \ref{fig4}c and \ref{fig4}d). The characteristic circular swimming time scale $\tau_s$, corresponding to the time for the bacterium to swim in a half circle, is associated with the start of the caging plateau in the tracer MSD. 
The cage breaking occurs over a time scale $\tau_c$ that the circular motion of the bacterium is disturbed by active noise. 
For a tracer interacting with multiple bacteria, we expect that $\tau_c$ will vary with changes in bacterial density and change in active noise.

To quantify the effects of active diffusion and swimmer density, we simulate trajectories of a single tracer advected by $N_s$ from 2 to 128 swimmers with \(D_{act}=0.2\) to \SI{6.4}{\micro m^2 s^{-1}}.
The diffusion coefficient of the tracers due to hydrodynamic interaction with bacteria, $D_h$, is obtained at long lag times where the MSD increases linearly.
Two different tracer diffusion processes are observed. 
For low $N_s$ or weak $D_{act}$, as suggested by the pairwise interaction between a single swimmer and tracer, the tracer MSD (for example, Fig. \ref{fig6}b) evolves from an initial super-diffusive regime through a sub-diffusive plateau indicating the caging effect of moving along looped trajectories, towards a diffusive regime. 
The decorrelation time scale at which the tracer becomes uncaged, $\tau_c$, decreases with increasing $N_s$ or $D_{act}$. 
Above some threshold values for these quantities, $\tau_c$ becomes comparable to $\tau_s$, and the form of the tracer MSD changes abruptly as the caging effect disappears. 
At such high $N_s$ or large $D_{act}$, the MSD evolves from superdiffusive at short lag time with $\alpha<2$ towards diffusive with no apparent caging at intermediate lag time, Fig. \ref{fig6}c. 
The $N_s$ and $D_{act}$ that yield caged behavior are summarized in Fig. \ref{fig6}d as square symbols and the cases for which tracers do not become caged are depicted by circles. 
The predicted effective hydrodynamic diffusivity $D_h$ at given $N_s$ and $D_{act}$, indicated by the symbol colors, increases sharply as the system transitions to uncaged states and increases thereafter with $\rho_s$. 
These simulation results show that increases in swimmer number $N_s$ and active diffusivity $D_{act}$ randomize tracer direction, weaken tracer caging, and enhance tracer diffusion. 
The $D_h$ simulated for $\rho_s>\SI{0.006}{\micro m^{-2}}$ is of the same order of magnitude as the Brownian diffusivity of micron sized colloids, indicating that hydrodynamic interactions can significantly impact tracer displacements. 
This effect can be amplified by increasing swimmer velocity, dipolar strength, and density.

\section{Interactions among multiple swimmers generate active noise}
\begin{figure*}
\centering
\includegraphics[height=5cm]{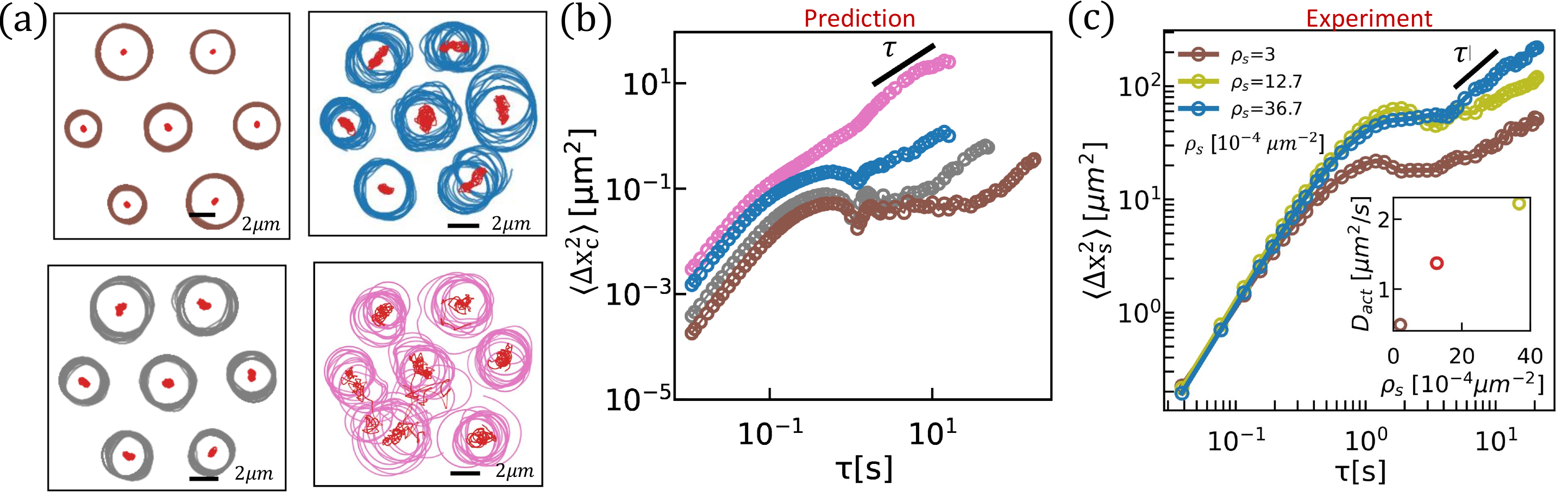}
\caption[Interactions among multiple swimmers generate enhanced diffusion]{\textbf{Interactions among multiple swimmers generate enhanced diffusion.} (a) Simulated trajectories of interacting swimmers with decrease in
their separation distance from \SI{8}{\micro m} shown in brown to \SI{6}{\micro m} shown in blue/pink. Radius of curvature is selected randomly between \SI{1.5}{\micro m} and \SI{2.5}{\micro m}. The center of rotation of each trajectory is
depicted in red dots. (b) MSD of the center of rotation of the ``center
bacterium'', $\langle \Delta x_c^2 \rangle$, graphed with the same color of trajectories shown in (a). (c)
Experimental MSD of PA01 at hexadecane aqueous interfaces at different
surface densities of swimmers. Inset: active diffusion coefficient of bacteria,
\(D_{act}\), extracted at large lag time. Each MSD is calculated from 12 swimmer trajectories.
}
\label{fig7}
\end{figure*}

Hydrodynamic interactions among swimmers at high surface  densities are a 
source of noise over lag times significant compared to the time to swim in circular trajectories. 
We simulate the pairwise interaction of
multiple pushers moving on circular paths with radii selected randomly
from \SI{1.5}{\micro m} to \SI{2.5}{\micro m} and with random initial orientations (Fig. \ref{fig7}a). The effect of reorientation from all neighboring swimmers is
superposed to update their swimming directions and superposed
displacements are used to update the swimmers' displacements. The
resulting MSD of the center of rotation of the ``center bacterium'',
shown in Fig. \ref{fig7}b, is reported to illustrate the effect of hydrodynamic
interactions as a function of surface density of neighboring swimmers.
The predicted MSD is ballistic at short lag time, caged thereafter and
diffusive at long lag time. 
Hydrodynamic interactions are highly
sensitive to the separation distance between nearest neighbors. The
timescale for the center of rotation to become diffusive decreases from
$\sim\SI{20}{ s}$ to $\sim\SI{1}{s}$ and the diffusivity increases
from \SI{0.0005}{\micro m^2 s^{-1}} to \SI{0.02}{\micro m^2 s^{-1}} as the center-to-center distance between
neighboring swimmers decreases from \SI{8}{\micro m} (brown trajectories in Fig. \ref{fig7}a)
to \SI{7}{\micro m} (grey trajectories in Fig. \ref{fig7}a)
and \SI{6}{\micro m} (blue and pink trajectories in Fig. \ref{fig7}a). 
As the bacteria become more densely packed, rare events in which two swimmers become proximate can
generate torques and forces with large magnitudes that destabilize the
swimmer paths as shown for the pink trajectories in Fig. \ref{fig7}a; the corresponding MSD of the center of rotation becomes purely diffusive as shown in pink in Fig. \ref{fig7}b. 
Such diffusion of the center of rotation will lead to diffusion in bacterial path at long lag times.
These simulations must be treated
with care, as they ignore higher order hydrodynamic modes and other
phenomena like flagellar entanglements that can occur for bacteria in
the very near field. Notably, similar to active-passive interactions
described in Fig. \ref{fig6}, fluctuations in bacterial speed,
angular speed, and the active noise in swimmer paths from other sources
can further randomize the paths of their neighboring swimmers resulting
in higher chance of near-field interactions. 
The randomizing effect of hydrodynamic
interactions can also lead to higher chances of bacterial collision and near-contact interactions.
Therefore, prior to the regime at high bacterial density where hydrodynamic interaction between swimmers generates coherence, at intermediate density these interactions promote chaos in the system.
While direct comparison of
experiment and prediction must be treated with care, active noise is indeed observed
for bacteria trapped at fluid interfaces. In experiment, we have
measured the MSDs of bacteria at interfaces with increasing surface
density \(\rho_{s}\) and extracted the diffusion coefficients at large
lag times, as reported in Fig. \ref{fig7}c. For all surface densities, the MSDs oscillate at moderate lag times reflecting their circular motion and become diffusive at long lag times. The orientation of bacteria at high surface
density decorrelates more rapidly and the MSDs show higher diffusivities.
These effects can be attributed in part to the predicted hydrodynamic
interactions in pusher mode, although our bacteria also switch between pusher and puller modes. 

\section{Conclusions}
We analyze the impact of interfacially-trapped bacteria on interfacial transport based on their induced flow fields. We calculate tracer displacements due to hydrodynamic interaction with swimmers and compare the predicted tracer paths to those constructed from experiment. 
We find that while hydrodynamic interaction with well-separated swimmers drives tracers in closed loops, net tracer displacement can occur for swimmers and tracers in closer proximity. 
These displacements rely on broken symmetry in swimmer-tracer separation, and lead to loopy tracer paths around the swimmer at long lag times. 
We also find that swimmers can attract or repel each other due to hydrodynamic interactions; such interactions among multiple swimmers are likely to induce active diffusion that has been observed for bacteria at interfaces. 
To further assess the effect of interfacial pushers on colloidal transport, we predict the interactions of a tracer particle with multiple swimmers with various active diffusivities. 
We find a “caging” effect in tracer mean square displacements owing to the looplike structure of tracer paths; this caging is eliminated by swimmer’s active diffusivity and by multiple swimmers-tracer interactions. 
The combined effects of active diffusion and multiple swimmer-tracer interactions generate a hydrodynamic diffusivity that can be comparable to thermal diffusivity.

In nature and in biotechnology, these results have implications in understanding the initial stages of biofilm formation at fluid interfaces, including cluster formation and nutrient supply near the bacteria-populated interface. Furthermore, these results can inform the design of biomimetic active colloids as active surface agents to promote interfacial mass transport to improve the efficiency of interfacial reaction and separation processes. 
One could envision design strategies for active colloids that impose particular trapping orientations and positions of propulsion sites to promote mixing that could significantly impact chemical engineering processes like reactive separations near fluid interfaces.

\textbf{Acknowledgements.}This research was made possible by a grant from the National Science Foundation (NSF grant no. CBET-1943394 and DMR-1607878)

\bibliography{main}
\end{document}